\documentclass[aps,11pt]{revtex4-1}

\newtheorem{theorem}{Theorem}
\newtheorem{proposition}{Proposition}
\newtheorem{lemma}{Lemma}
\newtheorem{corollary}{Corollary}

\usepackage{graphicx}
\usepackage{amsfonts}
\usepackage{amsmath}
\usepackage{tabulary}
\usepackage{mathrsfs}
\usepackage{eucal}
\usepackage{appendix}
\usepackage[ruled]{algorithm2e}
\usepackage{natbib}

\newcommand\bR{\mathbb{R}}
\newcommand\bE{\mathbb{E}}
\newcommand\bB{\mathbb{B}}
\newcommand\mO{\mathcal{O}}
\newcommand\mC{\mathcal{C}}
\newcommand\mG{\mathcal{G}}
\newcommand\mF{\mathcal{F}}

\newcommand\mP{\mathcal{P}}
\newcommand\mR{\mathcal{R}}

\newcommand\mL{\mathcal{L}}
\newcommand\mQ{\mathcal{Q}}

\newcommand\bC{\mathbf{C}}

\def\o{\omega}

\newcommand\qed{$\square$}

\begin{document}


\title{Towards the full information chain theory: solution methods for optimal information acquisition problem}
\author{E. Perevalov}
\email[E-mail: ]{eup2@lehigh.edu}
\author{D. Grace}
\email[E-mail: ]{dpg3@lehigh.edu}
\affiliation{Lehigh University\\ Bethlehem, PA}
\date{\today}

\begin{abstract}
When additional information sources are available in decision making problems that allow stochastic optimization formulations, an important question is how to optimally use the information the sources are capable of providing. A framework that relates information accuracy determined by the source's knowledge structure to its relevance determined by the problem being solved was proposed in a companion paper. There, the problem of optimal information acquisition was formulated as that of minimization of the expected loss of the solution subject to constraints dictated by the information source knowledge structure and depth. Approximate solution methods for this problem are developed making use of probability metrics method and its application for scenario reduction in stochastic optimization.
\end{abstract}

\pacs{02.50.Cw, 02.50.Le, 89.70.Cf}

\keywords{information; entropy; information theory; decision making}

\maketitle

\section{\label{s:intro}Introduction}
In many practically important decision making problems where uncertainty about input data is present and optimization methods are appropriate, sources of additional information are in principle available. Often, information that such sources possess fails to be taken advantage of due to its perceived and factual imprecision and to the lack of methodology that allows doing this in a regular controlled fashion. Such methodology, if developed in a general setting, would form a branch of the science of information which is at present represented by the classical Information Theory and its extensions. Generalizing somewhat, one can say that Information Theory explores the implications of information {\it quantity} while abstracting from the information content and, in particular, its {\it accuracy} and {\it relevance}. Respectively, the classical Information Theory is predominantly a theory of information transmission and related activities (such as compression). On the other hand, if information is to be acquired and used for decision making, a quantitative framework describing these processes would be rather helpful.

\begin{figure}
\centerline{\includegraphics*[scale=0.9]{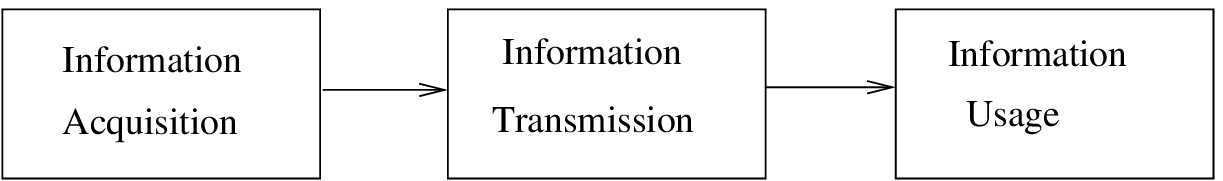}
} \caption{\label{f:Ichain}The full information chain.}
\end{figure}

A notion of {\it full information chain} was introduced in \cite{part1,part2} to schematically describe the typical path of information from acquisition to usage (see Fig.~\ref{f:Ichain} for an illustration). In this context, the classical Information Theory is a theory of the middle link, while the methodology developed in \cite{part1,part2,loss-part1} and the present article concerns the basics of a general theory of the two end links. More specifically, the information acquisition link was addressed in \cite{part1,part2}, and the basic framework for the information use link was proposed in \cite{loss-part1} resulting -- after making a connection with the results of \cite{part1,part2} -- in a formulation of the optimal information acquisition problem. This article builds on these results and proposes specific solution methods for the optimal information acquisition problem.

The proposed approach, as was mentioned earlier, can be looked upon as an attempt to initiate a process of extending the classical Information Theory to a theory of the whole information chain. The field of Information Theory, born from Shannon's work on the theory of communications \cite{SHANNON:1948}, since have enjoyed great success in a number of fields that include, besides communication theory, statistical physics \cite{JAYNES:1957a,JAYNES:1957b}, computer vision \cite{viola1995}, climatology \cite{mokhov2006,verdes2005},
physiology \cite{katura2006} and neurophysiology \cite{chavez2003}.  Generalized Information Theory (see e.g. \cite{klir1996},  \cite{maeda1993,harmanec1994})
addresses problems of characterizing uncertainty in frameworks that are more general than
classical probability such as Dempster-Shafer theory \cite{shafer1976}.

The approach developed here is based on a theory of information exchange between the agent and information source(s) that is developed in \cite{part1,part2}. The latter can be thought of as a development of the general theory of inquiry that goes back to the work of Cox \cite{cox1946,cox1961,cox1979}. This line of work received more attention recently resulting in a formulation of the calculus of inquiry \cite{knuth05,knuth07,knuth08}. The definition of questions adapted in \cite{part1} corresponds to the particular subclass of questions -- the partition questions -- defined in \cite{knuth05}. It is also related to the measure-dependent notion of a question introduced in \cite{caticha04}. Our work in \cite{part1,part2} goes beyond that on the calculus of inquiry in that it introduces the concept of {\it pseudoenergy} as a measure of source specific difficulty of various questions to the given information source. One could say that it develops a quantitative theory of {\it knowledge} as opposed to the theory of information.

One of successful applications of the order-theoretic approach to fundamental physics is the recent derivation \cite{knuth-sr} of Lorentz transformations and Minkowski metric of special relativity directly from the consideration of the partial order of events in space-time.

Information Physics \cite{caticha-rev} is a relatively new branch of physical sciences that studies the role information plays in fundamental laws of nature. This line of research goes back to the defining work of Jaynes \cite{JAYNES:1957a,JAYNES:1957b} on the application of the Principle of Maximum Entropy (MaxEnt) to derive the fundamental laws of thermodynamics. It is related to the proposed framework in that it addresses information {\it relevance} in application to physical sciences. The main Information Physics hypothesis is that the laws of nature are essentially the laws of inductive inference correctly applied to respective systems. In order to correctly formulate them one needs to know the degrees of freedom and the relevant information necessary to completely specify the system state. Recently, this approach (in modified and extended form) was applied to derive the fundamental laws of classical \cite{caticha07} and quantum \cite{caticha11} mechanics, and also -- very recently -- relativistic quantum theory \cite{caticha12}. A closely related line of research explores the ramifications of general order-theoretic relations. An interesting example of the latter is the derivation \cite{knuth-sr} of Lorentz transformations and Minkowski metric of special relativity directly from the consideration of the partial order of events in space-time.

The area of statistical decision making has dealt with the idea of improving solution quality by means of acquiring additional information. There have been applications to innovation
adoption \cite{mccardle1985}, \cite{jensen1988}, fashion decisions \cite{fisher1996} and vaccine composition decisions for flu immunization \cite{kornish2008} can be mentioned in this regard. Some
authors \cite{fischer1996}, \cite{ellison1993}  even introduced models
(e.g. effective information model) for accounting for the actual, or effective, amount of
information  contained in the received observations.  One could also mention the recent work on  optimal decision making in the absence of the knowledge of the distribution shape and parameters \cite{huh2009,levi2007,bassamboo2009}. The difference of the proposed approach is in that it explicitly describes and allows to optimize over not just the quantity of additional information but also its content and is based on explicit description of properties of information sources.

The related problem of optimal usage of information obtained from experts has been addressed in existing research literature mostly in the form of updating the agent's beliefs given probability assessment from multiple experts \cite{french1985,genest1986,clemen1987,clemen1999} and optimal combining of expert opinions, including experts with incoherent and missing outputs \cite{predd2008}. In the present and preceding papers, the emphasis is on {\it optimizing} on the particular type of information for the given expert(s) and decision making problem.

Methodologically, the present article borrows from the field of probability metrics and scenario reduction in stochastic optimization. More details, along with relevant references, can be found in Appendices.

The rest of the paper is organized as follows. In Section~\ref{s:1link}, main results of \cite{part1,part2} that are necessary for the developments in this paper are reviewed. Section~\ref{s:3link} reviews main results of \cite{loss-part1} where, in particular, the problem of additional information acquisition was formulated in the specific form that is used here. Section~\ref{s:method} develops the main theoretical framework for the use of scenario reduction methods for optimization of additional information acquisition. Section~\ref{s:example} provides an example illustrating the use of methods developed in Section~\ref{s:method}. Section~\ref{s:conclusion} contains a conclusion. Appendix~\ref{a:proofs} provides proofs for some of the results in the main text, Appendix~\ref{a:metrics} gives some background information on probability metrics in application to stochastic optimization, and Appendix~\ref{a:s-reduction} contains a very brief review of scenario reduction algorithms.

\section{\label{s:1link}Information Accuracy: Source Knowledge Structure}
As was explained in \cite{loss-part1}, the starting point of the whole discussion
is a problem of the general form
\begin{equation}
 \label{eq:gen_stoch}
 \mbox{min}_{x\in X} \int_{\Omega} f(\omega, x) P(d\omega).
 \end{equation}
 where $X$ is the set of all feasible solutions,  $\Omega$ is a parameter space to which uncertain problem parameters belong, and  $P$ is a fixed initial probability measure (with a suitable sigma-algebra assumed) on $\Omega$ that describes the initial state of the uncertainty. The function $f$: $\Omega\times X\rightarrow \overline \bR$ is assumed to be integrable on $\Omega$ for each $x\in X$. For example, in the context of stochastic optimization, $X$ is the set of feasible first-stage solutions and $f(\omega,x)$ is the best possible objective value for the first stage decision $x$ in case when the random outcome $\o$ is observed.

Let $L(P)$ be the expected loss corresponding to measure $P$ defined as follows.
\begin{equation}
L(P)=\int_{\Omega} f(\o,x^*_P) P(d\o)-\int_{\Omega} f(\o,x^*_{\o}) P(d\o),
\label{eq:L(P)}
\end{equation}
where $x^*_P$ is a solution of (\ref{eq:gen_stoch}) and $x^*_{\o}$ is a solution of $\mbox{min}_{x\in X} f(\o,x)$ for the given $\o$.

The main goal is, as explained in \cite{loss-part1}, for the given information source(s), to find the way of extracting information from it so that the resulting expected loss is minimized. The knowledge structure of the source determines the accuracy of source's answers. The difference between the original loss (\ref{eq:L(P)}) and the loss obtained with the help of the source's answers serves as a measure of the information relevance.

The process of information exchange between an agent and an information source was described in \cite{part1,part2}. Here we review the main results to make the presentation self-contained. The source is characterized by its {\it knowledge structure} encoded in the form the question difficulty functional described below. The agent poses questions for which the source provides answers.

Questions were identified in \cite{part1} with partitions $\bC=\{C_1,\dotsc, C_r\}$ of the parameter space $\Omega$ of the problem. Partitions were allowed to be incomplete, i.e. such that $\cup_{j=1}^r C_j\subset \Omega$. The {\it question difficulty} functional was introduced to measure the degree of difficulty of the question to the given information source, so that the information source would be able to answer questions with lower values of the difficulty functional more accurately that those with higher values of difficulty. The specific form of the difficulty functional was determined in \cite{part1} by demanding that it satisfy a system of reasonable postulates that, in particular, imposed the requirements of linearity and isotropy. The resulting form of the difficulty functional is given in the following theorem.

\begin{theorem}
Let the function $G(\Omega, \bC, P)$ where $\bC=\{C_1,\dotsc, C_r\}$ satisfy Postulates 1 through 6 (see \cite{part1}). Then it has the form
$$G(\Omega, \bC, P)=\frac{\sum_{j=1}^r u(C_j)P(C_j)\log \frac{1}{P(C_j)}}{\sum_{j=1}^r P(C_j)},$$
where $u(C_j)=\frac{\int_{C_j}u(\o)\,dP(\o)}{P(C_j)}$ and $u$: $\Omega\rightarrow \bR$  is an integrable nonnegative function on the parameter space~$\Omega$.
\label{th:G}
\end{theorem}

One can see that the difficulty of the given question $\bC$ depends on, besides the initial probability measure $P$, the function $u$ : $\Omega\rightarrow \bR_+$ on the parameter space $\Omega$.  Using parallels with thermodynamics (see \cite{part1} for more details), this function may be called the {\it pseudotemperature}. The question difficulty then can be interpreted as the amount of {\it pseudoenergy} associated with question~$\bC$.

Given a question $\bC=\{C_1,\dotsc, C_r\}$, the information source can provide an answer $V(\bC)$ that takes one of values in the set $\{s_1,\dotsc, s_m\}$. A reception of the value $s_k$ has an effect of modifying the original probability measure $P$ on $\Omega$ to a new (updated) measure $P^k$. To ensure the the answer $V(\bC)$ is in fact an answer to the (complete) question $\bC$ (and no more) the following condition is required to hold for the updated measures $P^k$, $k=1,\dotsc, m$:
\begin{equation}
P^k=\sum_{j=1}^r p_{kj}P_{C_j},
\label{eq:Pk-mc}
\end{equation}
where $p_{kj}$, $k=1,\dotsc, m$, $j=1,\dotsc, r$ are nonnegative coefficients such that $\sum_{j=1}^r p_{kj}=1$ for $k=1,\dotsc, m$.

The {\it answer depth} functional $Y(\Omega,\bC,P, V(\bC))$ for the answer $V(\bC)$ to question $\bC$ measures the amount of {\it pseudoenergy} that is conveyed by $V(\bC)$ in response to question $\bC$. The general form of $Y(\Omega,\bC,P, V(\bC))$ can be established if certain reasonable requirements (postulates) it has to satisfy are imposed. A system of postulates proposed in \cite{part2} that parallels the postulates for question difficulty and, in particular, imposes the requirements of linearity and isotropy. The following theorem was then proved in \cite{part2}.

\begin{theorem}
The answer depth functional $Y(\Omega, \bC, P, V(\bC))$ has the form
$$Y(\Omega, \bC, P, V(\bC))=\sum_{k=1}^m \Pr(V(\bC)=s_k)\frac{\sum_{j=1}^r u(C_j)P^k(C_j)\log \frac{P^k(C_j)}{P(C_j)}}{\sum_{j=1}^r P^k(C_j)}, $$
where $P^k$ is the measure on $\Omega$ updates by the reception of $V(\bC)=s_k$ and
$u(C_j)=\frac{1}{P(C_j)}\int_{C_j} u(\o) dP(\o)$  and the function $u$: $\Omega\rightarrow \bR$ is the same function that is used in the question difficulty functional $G(\Omega, \bC,P)$.
\label{th:Y}
\end{theorem}

It can be shown (see \cite{part2} for details) that if $V(\bC)$ is any answer to the question $\bC$ then $Y(\Omega, \bC, P, V(\bC))\le G(\Omega, \bC, P)$  with equality if and only if the answer $V(\bC)=V^*(\bC)$ is {\it perfect}, i.e. $P^j=P_{C_j}$ for $j=1,\dotsc, r$.

As far as answers that are not perfect are concerned, it is convenient to consider the class of answers for which the degree of imperfection is described by a single error probability $\alpha$ -- the {\it quasi-perfect} answers \cite{part2}. For a quasi-perfect answer $V_{\alpha}(\bC)$ to a (complete) question $\bC=\{C_1,\dotsc, C_r\}$, the coefficients $p_{kj}$ have the form $p_{kj}=(1-\alpha)\delta_{k,j}+\alpha P(C_j)$ for $k=1,\dotsc, r$ and $j=1,\dotsc, r$, and the updated measure $P^k$ is simply
 \begin{equation*}
 P^k = \alpha P + (1-\alpha) P_{C_k}.
 \end{equation*}
 for $k=1,\dotsc, r$.
 Clearly, for $\alpha=0$ a quasi-perfect answer to $\bC$ becomes a perfect one. It can be shown (see \cite{part2}) that the answer depth functional for a quasi-perfect answer $V_{\alpha}(\bC)$ to question $\bC$ can be written as
 \begin{equation*}
 \begin{split}
 Y(\Omega,\bC,P,V_{\alpha}(\bC))&=\sum_{k=1}^r u(C_k)P(C_k)(1-\alpha +\alpha P(C_k))\log \frac{1-\alpha +\alpha P(C_k)}{P(C_k)}\\
 &+ \alpha \log \alpha \sum_{k=1}^r u(C_k)P(C_k)(1-P(C_k)),
 \end{split}
 \end{equation*}
 which can be seen to reduce to $G(\Omega,\bC,P)$ for $\alpha=0$ (when $V(\bC)=V^*(\bC)$) and vanish for $\alpha=1$.

An information source model provides a connection between questions and the corresponding answers. It was defined in \cite{part2} as a function $h$~:~$\bR_+\rightarrow \bR_+$ such that
$$Y(\Omega,\bC,P,V(\bC))=h(G(\Omega,\bC,P)).$$

The simplest information source model considered in \cite{part2} is the {\it simple capacity model} given by
\begin{equation}
h(x)=\begin{cases} x & \mbox{if}\; x\le Y_s \\
                   Y_s & \mbox{if}\; x>Y_s.
\end{cases}
\label{eq:Y(G)-cap-simple}
\end{equation}
which is fully characterized by the single parameter $Y_s$ which has the meaning of the information source capacity. The most apparent drawback of model (\ref{eq:Y(G)-cap-simple}) is that, according to it, the source  provides a perfect answer to any question with difficulty not exceeding the source capacity. The {\it linear modified capacity model} described by
\begin{equation}
h(x)=\begin{cases} bx & \mbox{if}\; x\le \frac{Y_s}{b} \\
                   Y_s & \mbox{if}\; x>\frac{Y_s}{b}
\end{cases}
\label{eq:Y(G)-cap-mod}
\end{equation}
removes this drawback at the expense of one extra parameter $b\le 1$ that has to be estimated. Several other models were proposed in \cite{part2}.

The values of model parameters as well as pseudotemperature functions for information sources can be estimated from the observed sources' performance on sample questions. The corresponding estimation procedures were also discussed in \cite{part2}.

\section{\label{s:3link}Information Relevance: Loss Reduction}
The basic framework for the information use link description was presented in \cite{loss-part1}. We briefly summarizes the main points here to make a transition to the subject of this article in a self-contained manner.

 Consider the set $\mG$ of all maps from $\Omega$ into $X$ with a discrete image set. Any such map $g\in \mG$ can be uniquely described by the corresponding partition $\bC=\{C_1,\dotsc, C_r \}$ of $\Omega$ and the corresponding image set $I=\{x_1,\dotsc, x_r\}$ such that $g(\o)=x_j$ for all $\o\in C_j$.
Let $P$ be any probability measure on $\Omega$, let $x$ an arbitrary element of the solution space $X$, and let  $g\in \mG$ be an arbitrary map from $\Omega$ into $X$. The {\it suboptimality}, {\it loss} and {\it gain} functionals are defined (\cite{loss-part1}) as follows.
\begin{equation}
\label{eq:subopt}
S(x,P)=\bE_P f(\omega, x)-\bE_P f(\omega, x_P^*)=\int_{\Omega} (f(\omega, x)-f(\omega, x_P^*)) P(d\omega),
\end{equation}
\begin{equation}
\label{eq:loss}
L(g,P) = \bE_P f(\omega, g(\omega)) - \bE_P f(\omega, x_{\omega}^*)= \int_{\Omega} (f(\omega, g(\omega))-f(\omega, x_{\omega}^*)) P(d\omega),
\end{equation}
and
\begin{equation}
\label{eq:gain}
B(g,P)=\bE_P f(\omega, x_P^*)- \bE_P f(\omega, g(\omega)) = \int_{\Omega} (f(\omega, x_P^*)-f(\omega, g(\omega))) P(d\omega).
\end{equation}
respectively.

Moreover, it is convenient to introduce the corresponding functionals not just for a fixed measure $P$, but also for the given question $\bC$ and a given answer $V(\bC)$, For example,
for an arbitrary $x\in X$, the suboptimality of solution $x$ with respect to question $\bC$ (and initial measure $P$) is given by
\begin{equation}
S(x,P_{\bC})=\sum_{i=1}^s P(C_j) S(x,P_{C_j}),
\label{eq:S(bC)}
\end{equation}
and the suboptimality of $x$ with respect to answer $V(\bC)$ to question $\bC$ (and initial measure $P$) reads
\begin{equation}
S(x,P_{V(\bC)})=\sum_{k=1}^m v_k S(x,P^k),
\label{eq:S(V)}
\end{equation}
where $v_k\equiv \Pr(V(\bC)=s_k)$ for brevity.
The loss and gain functionals for the given map $g\in \mG$ and question $\bC$ and answer $V(\bC)$ are defined analogously.

Note that each map $g=(\bC(g),I(g))$ from the set $\mG$ can be characterized by the corresponding loss $L(g,P)$ with respect to the original measure $P$ and the value $G(\Omega, \bC(g), P)$ -- the difficulty of the corresponding question. The {\it efficient frontier} in the Euclidean plane with coordinates $(G(\Omega, \bC(g), P), L(g,P))$ can be found by solving the following parametric optimization problem
\begin{equation}
\begin{aligned}
& \underset{g\in \mG}{\text{minimize}}
& & L(g,P)  \\
& \text{subject to}
& & G(\Omega, \bC(g),P)\le \gamma \\
\end{aligned}
\label{eq:Pareto}
\end{equation}
for all values of the parameter $\gamma$.

The maps $g$ that are solutions of (\ref{eq:Pareto}) for various values of the parameter $\gamma$ possess the property of having the smallest possible loss among the maps corresponding to questions whose difficulty does not exceed the given value $\gamma$. We denote by $\mO$ the subset of all maps in $\mG$ that are solutions of (\ref{eq:Pareto}) and by $\mC$ the set of all {\it subset-optimal} maps, i.e. maps of the form $(\{C_1, \dotsc, C_r\}, \{x^*_{P_{C_1}}, \dotsc, x^*_{P_{C_r}}\})$, where $x_{P_{C_j}}^*$ is an optimal solution of problem (\ref{eq:gen_stoch}) with measure $P$ replaced with the conditional measure $P_{C_j}$. Then, as was shown in \cite{loss-part1},
\begin{equation}
\mO\subseteq \mC,
\label{eq:OinC}
\end{equation}
i.e. if one is interested in finding Pareto-optimal maps in $\mO$ it is sufficient to consider subset-optimal maps only. We call a partition $\bC$ {\it optimal} if the corresponding map $g=(\{C_1,\dotsc,C_r\},\{x_{P_1}^*,\dotsc, x_{P_r}^*\})\in \mC$ belongs to the set $\mO$ of Pareto-optimal maps. So the problem of finding maps in the set $\mO$ is equivalent to that of searching for optimal partitions of the parameter space $\Omega$.

Let us now address the optimal information acquisition problem (\ref{eq:minLV}): what question(s) need to be asked the given information source in order to obtain the minimum possible loss for (\ref{eq:gen_stoch}). Given a question $\bC=\{C_1,\dotsc, C_r\}$ to an information source and its answer $V(\bC)$ taking values in the set $\{s_1,\dotsc, s_m\}$, we denote
by $\mL(s_k)$, $k=1,\dotsc, m$ the {\it minimum conditional expected loss} given that $V(\bC)=s_k$ and by  $\mL(V(\bC))$ the {\it minimum expected loss} that the agent can achieve given the answer $V(\bC)$. The latter can be found as
\begin{equation}
\mL(V(\bC))=\sum_{k=1}^m \Pr(V(\bC)=s_k)\mL(s_k),
\label{eq:mLV}
\end{equation}
i.e. as an expectation over possible values of the answer $V(\bC)$.

If the agent poses a question $\bC=\{C_1,\dotsc, C_r\}$ to the information source and receives a particular value $s_k$ of answer $V(\bC)$, the original measure $P$ on $\Omega$ gets updated to $P^k$. Therefore, in order to minimize loss for the given value $s_k$ of answer $V(\bC)$, the agent needs to choose the solution $x^*_{P^k}$ -- the solution minimizing the expectation $\bE_{P^k} f(\o,x)$ over all (feasible) values of $x$.

The next two propositions, proved in \cite{loss-part1}, give the minimum expected loss achievable with a perfect and a general answer to question $\bC$, respectively.

\begin{proposition}
Let $\bC=\{C_1,\dotsc, C_r\}$ be a complete question and $g_{\bC,P}\in \mC$ be a corresponding subset optimal map. If the agent is given a perfect answer $V^*(\bC)$ to $\bC$ then
$$\mL(V^*(\bC))=L(g_{\bC,P},P). $$
\label{p:loss-perf}
\end{proposition}

\begin{proposition}
Let $\bC=\{C_1,\dotsc, C_r\}$ be a complete question and $g_{\bC,P}\in \mC$ be a corresponding subset optimal map. If the agent is given a (generally imperfect) answer $V(\bC)$ to $\bC$ then
$$\mL(V(\bC))=B(g_{\bC,P},P_{V(\bC)})+L(g_{\bC,P},P). $$
\label{p:loss-imp}
\end{proposition}

The information acquisition optimization problem can then be written as
\begin{equation}
\begin{aligned}
& \underset{\bC}{\text{minimize}}
& & \mL(V(\bC))  \\
& \text{subject to}
& & Y(\Omega,\bC,P,V(\bC))=h(G(\Omega,\bC,P)),\\
\end{aligned}
\label{eq:minLV}
\end{equation}
where the minimum expected loss $\mL(V(\bC))$ is given by either Proposition~\ref{p:loss-perf} or Proposition~\ref{p:loss-imp}. The source model function $h(\cdot)$ and the pseudotemperature function $u(\cdot)$ that enters the expressions for the question difficulty and answer depth in (\ref{eq:minLV}) are assumed to be known.

It's easy to see that if a source is capable of providing perfect answers (for instance, in the simple linear model) solution of problem (\ref{eq:minLV}) reduces to finding the efficient frontier: if $L^*(G)$ is the expression describing the efficient frontier (abstracting from its true discrete structure) and $Y_s$ is the capacity of the information source, then the minimum in (\ref{eq:minLV}) is equal to $L^*(Y_s)$ and is achieved by the question $\bC$ lying on the efficient frontier such that $G(\Omega,\bC,P)=Y_s$.

If a source cannot provide perfect answers, questions with difficulty exceeding the source capacity need to be considered in order to minimize the expected loss. The search for an optimal question in this case becomes more complicated as the error structure for the source's answers needs to be taken into account. If answers are assumed to be quasi-perfect, optimal question(s) can be found approximately provided the efficient frontier is already known.

\section{\label{s:method}Information Acquisition Optimization}
As stated in the previous section, a solution of the optimal information acquisition problem (\ref{eq:minLV}) is greatly facilitated by the search for the efficient frontier $L^*(G)$ in the set $\mG$ of all questions. To find the latter, one needs to determine optimal partitions of the parameter space $\Omega$. It turns out that the methods of measure (scenario) reduction developed previously for solving stochastic optimization problems can be also helpful for the task of searching for optimal partitions.

\subsection{Measure reduction and optimal partitions}
In the following, we assume that the (initial) probability measure $P$ is supported at a discrete set $\{\o_1,\dotsc, \o_N\}\equiv \Omega_N\subset \Omega$:
\begin{equation}
P=\sum_{i=1}^N p_i \delta_{\o_i},
\label{eq:Pd}
\end{equation}
where $\delta_{\o}$ is a Dirac delta that puts a unit mass at $\o$. Points $\o_i\in \Omega_N$ are usually referred to as {\it scenarios}. The {\it scenario reduction} methodology (see~Appendix C) is often used in stochastic optimization to lower computational complexity of various practically important problems. In scenario reduction approach, the original discrete measure $P$ given by (\ref{eq:Pd}) is said to be {\it reduced} to another discrete measure $Q$ given by
\begin{equation}
Q=\sum_{j=1}^M q_j \delta_{\tilde\o_j},
\label{eq:Qd}
\end{equation}
if the support $\{\tilde \o_1,\dotsc, \tilde \o_M\}$ of $Q$ is a subset of $\Omega_N$.

For later convenience, we denote by $\mR_M(\Omega_N)$ the set of all {\it scenario reduction maps} from the set of measures of the form (\ref{eq:Pd}) supported at $\Omega_N$ into the set of all measures of the form (\ref{eq:Qd}) supported at some subset of $\Omega_N$ of cardinality $M<N$ satisfying the additional property that we call {\it simplicity}. A map $\nu\in \mR_M(\Omega_N)$ is called {\it simple} if there exists a partition $\{S_1,\dotsc, S_M\}$ of the set of scenarios $\Omega_N$  such that  $\nu(\o_i)=\tilde \o_j$ for all $\o_i\in S_j$ and $q_j=\sum_{\{i: \o_i\in S_j\}} p_i$. In such a case we write $Q=\nu(P)$ and $S_j=\nu^{-1}(\tilde \o_j)$ for $j=1,\dotsc, M$.

Additionally, if $c$: $\Omega\times\Omega\rightarrow \bR_+$ is some symmetric cost function, we call a map $\nu\in \mR_M(\Omega_N)$ {\it $c$-optimal} if $i=\arg\min c(\o_i,\nu(\o_i))$ for $i=1,\dotsc, N$. It is shown in \cite{heitsch2003} that the Monge-Kantorovich functional (see~Appendix B) $\hat\mu_c(P,Q)$ is minimized for all measures $Q$ supported at $\{\tilde\o_1,\dotsc, \tilde\o_M\}=\nu(\Omega_N)$ iff the corresponding simple scenario reduction map is $c$-optimal.

In the following we call measures $P$ and $Q$ $\bC$-{\it equivalent} for some partition $\bC$ of $\Omega$ if $P(C)=Q(C)$ for all $C\in \bC$. It is easy to see that measures $P$ and $Q$ are $\bC$-equivalent for all possible partitions $\bC$ if and only if $P=Q$, but two distinct measures can easily be $\bC$-equivalent for a specific partition $\bC$. In particular, any two measures on $\Omega$ are $\bC$-equivalent if $\bC$ is the trivial partition $\bC=\{\Omega\}$.

Given a probability measure $P$ on $\Omega$ and some measure $Q$ that was obtained from $P$ by a reduction, let us denote by $\mQ(Q|P)$ the {\it virtual pseudoenergy} content of measure $Q$ relative to $P$. It is defined as follows.
\begin{equation}
\mQ(Q|P)=G(\Omega,\bC_f(P),P)-G(\Omega,\bC_f(Q),Q),
\label{eq:mQ}
\end{equation}
i.e. $\mQ(Q|P)$ is the difference between the difficulties of exhaustive questions  associated with measures $P$ and $Q$, respectively. One can think about the virtual pseudoenergy of $Q$ relative to $P$ as an amount of pseudoenergy a source would need to supply in order to obtain a new state in which the hardest possible question has a difficulty equal to $G(\Omega,\bC_f(Q),Q)$. Since no question is in fact answered in going from measure $P$ to the reduced measure $Q$, we call this pseudoenergy virtual.

We can now introduce the {\it virtual difficulty} of question $\bC$  for measure $Q$ with respect to measure~$P$:
\begin{equation}
G_P(\Omega, \bC,Q)=\mQ(Q|P)+ G(\Omega, \bC,Q).
\label{eq:G_P}
\end{equation}
In particular, $G_P(\Omega, \bC,P)=G(\Omega, \bC,P)$, i.e. the virtual difficulty of $\bC$ for measure $P$ relative to $P$ reduces just to the standard difficulty of $\bC$.

 It also turns out to be useful to introduce the {\it relative expected loss} for partitions of $\Omega$ and measures $Q$ obtained from the original measure $P$ by a (simple) scenario reduction operation. In other words, we assume that there exists $\nu\in \mR_M(\Omega_N)$ for some value of $M<N$ such that $Q=\nu(P)$. The relative (to measure $P$) expected loss of partition $\bC$ and measure $Q$ is then defined as follows.
\begin{equation}
L_P(\bC,Q)=\sum_{C\in \bC} P(C)L(g_{\bC,Q},P),
\label{eq:L_P}
\end{equation}
where $g_{\bC,Q}$ is the subset-optimal map for partition $\bC$ and measure $Q$. In particular, if $\bC$ is the trivial partition $\bC=\{\Omega\}$, the loss of $Q$ relative to $P$ is simply\footnote{Here and later we omit the trivial partition from the list of arguments of $G(\cdot)$ and $L(\cdot)$.} $L_P(Q)=L(g_Q,P)$. If the measure $Q$ coincides with $P$, the loss relative to $P$ is just the standard expected loss of the corresponding subset-optimal map: $L_P(\bC,P)=L(g_{\bC,P},P)$.

Let us now consider the following construction.
Reduce the original measure $P$ to $Q$ that is supported at $r$ points: $Q=\nu(P)$, where $\nu\in \mR_r(\Omega_N)$. Let $Q=\sum_{j=1}^r q_j\tilde \o_j$ and let $S_j$ the preimage of $\tilde \o_j$ under map $\nu$: $\nu(\o_i)=\tilde \o_j$ for all $\o_i\in S_j$. Then let $\bC$ be a partition of $\Omega$ such that $S_j\subset C_j$ for $j=1,\dotsc, r$. We say that the partition $\bC$ is {\it generated} by the map $\nu\in \mR_r(\Omega_N)$, or, equivalently by the reduction of measure $P$ to $Q$. Let $\hat\bC$ be an arbitrary coarsening of $\bC$.

We are interested in the location of points $P$, $Q$, $(\bC,P)$, $(\bC,Q)$, $(\hat\bC,P)$ and $(\hat\bC,Q)$ on the plane with coordinates $(G_P(\Omega,\cdot), L_P(\cdot))$. First of all, it is clear that $G_P(\Omega,P)=0$ and $L_P(P)=L(g_{P},P)$ where $L(g_P,P)$ is the EVPI of problem (\ref{eq:gen_stoch}). Second, it is also clear that
\begin{equation}
\label{eq:GP(CQ)}
\begin{split}
G_P(\Omega,\bC,Q)&=\mQ(Q|P)+G(\Omega,\bC,Q)\\
  &=G(\Omega,\bC_f(P),P)-G(\Omega,\bC_f(Q),Q)+G(\Omega,\bC,Q)\\
&=G(\Omega,\bC_f(P),P)
\end{split}
\end{equation}
since $\bC=\bC_f(Q)$ by construction of $Q$. In words, the virtual difficulty of the question $\bC$ for measure $Q$ where the partition $\bC$ was generated by a reduction of the original  measure $P$ to $Q$ is equal to the difficulty of the exhaustive question for the original measure $P$.

To obtain relationships between relative expected losses the following two auxiliary lemmas are needed.
\begin{lemma}
Let $c_{ij}=c_{ji}$, $i,j=1,\ldots, N$ be a symmetric matrix with elements $c_{ij}$ satisfying the triangle inequality $c_{ij}\le c_{ik}+c_{kj}$. Let $\{p_i\}_{i=1}^N$ be a probability distribution. Then
$$\sum_{i=1}^N \sum_{j=1}^N p_jp_j c_{ij}\le 2\min_{i} \sum_{j=1}^N p_jc_{ij}. $$
\label{l:av<2min}
\end{lemma}

{\bf Proof:} See Appendix A. \qed

The second lemma states a useful probability metrics result. Let $P=\sum_{i=1}^N p_i\delta_{\o_i}$ be a discrete support probability measure on $\Omega$ and let $Q=\sum_{i=1}^M q_i\delta_{\tilde\o_i}$ be another such measure. Let $\zeta_c(P,Q)$ be a Fortet-Mourier metric for some cost function $c$ : $\Omega\times\Omega\rightarrow\bR_+$ that satisfies conditions described in~Appendix B. Finally, let $\bC=\{C_1,\dotsc, C_r\}$ be a partition of $\Omega$ such that the measures $P$ and $Q$ are $\bC$-equivalent.
\begin{lemma}
Under assumptions described above,

1. $\zeta_c(P,Q)\le \sum_{j=1}^r w_j\zeta_c(P_{C_j},Q_{C_j})$, where $w_j=P(C_j)=Q(C_j)$.

2. If $Q$ is generated by some map $\nu\in \mR_r(\Omega_N)$ that is $\hat c$-optimal, where $\hat c$ is the reduced cost function defined as in (\ref{eq:reduced-c}) then

\hspace{3mm} $\zeta_c(P,Q)= \sum_{j=1}^r w_j\zeta_c(P_{C_j},Q_{C_j})$.
\label{l:zeta_c}
\end{lemma}

{\bf Proof:} See Appendix A. \qed

Now, assume that the integrand $f(\o,x)$ in (\ref{eq:gen_stoch}) is in class $\mF_c$ defined in~Appendix B for some symmetric cost function $c$ : $\Omega\times\Omega\rightarrow \bR_+$ that satisfies the conditions described in~Appendix B.
The following proposition describes a relation between relative expected losses for measures $P$ and~$Q$.
\begin{proposition}
Let $\bC$ be a partition of $\Omega$ generated by a reduction of a measure $P$ with support at $\Omega_N\subset \Omega$ to $Q$ by means of a $\hat c$-optimal map $\nu\in \mR_r(\Omega_N)$ and let $\hat\bC$ any coarsening of $\bC$ (including $\bC$ itself). Then
$$L_P(\hat\bC,Q)\le L_P(\hat\bC, P) + 2K\zeta_c(P,Q),$$
where $K>0$ is some constant that does not depend on measures $Q$ and $P$.
\label{p:L<L+mu}
\end{proposition}

{\bf Proof:} See Appendix A. \qed

If we use the trivial partition $\hat\bC=\{\Omega\}$ (which is obviously a coarsening of any $\bC$) in Proposition~\ref{p:L<L+mu} we can obtain an upper bound on the relative loss of $Q$ with respect to $P$ which we formulate as a corollary.
\begin{corollary}
The loss of reduced measure $Q$ relative to $P$ can be bounded from above as
$$L_P(Q)\le L(g_P,P)+2K\zeta_c(P,Q), $$
where $L(g_P,P)\equiv L_P(P)$ is the EVPI of the original problem (\ref{eq:gen_stoch}).
\label{c:relQP}
\end{corollary}

The following proposition relates the expected loss of a subset-optimal map based on a partition generated by a reduction of the original measure $P$ to measure $Q$ to the Fortet-Mourier distance between $P$ and $Q$.
\begin{proposition}
Let $\bC$ be a partition of $\Omega$ generated by a reduction of a measure $P$ supported at the discrete set $\Omega_N\subset \Omega$ to measure $Q$ by means of a $\hat c$-optimal map $\nu\in \mF_r(\Omega_N)$. Then
$$L_P(\bC,P)\equiv L(g_{\bC,P},P)\le 2K\zeta_c(P,Q),$$
where $K>0$ is a constant.
\label{p:L<mu}
\end{proposition}

{\bf Proof:} See Appendix A. \qed

Fig.~\ref{f:lossub} shows the locations of various points on $(G_P(\Omega,\cdot,\cdot), L_P(\cdot,\cdot))$ coordinate plane.

\begin{figure}
\centerline{\includegraphics*[scale=0.7]{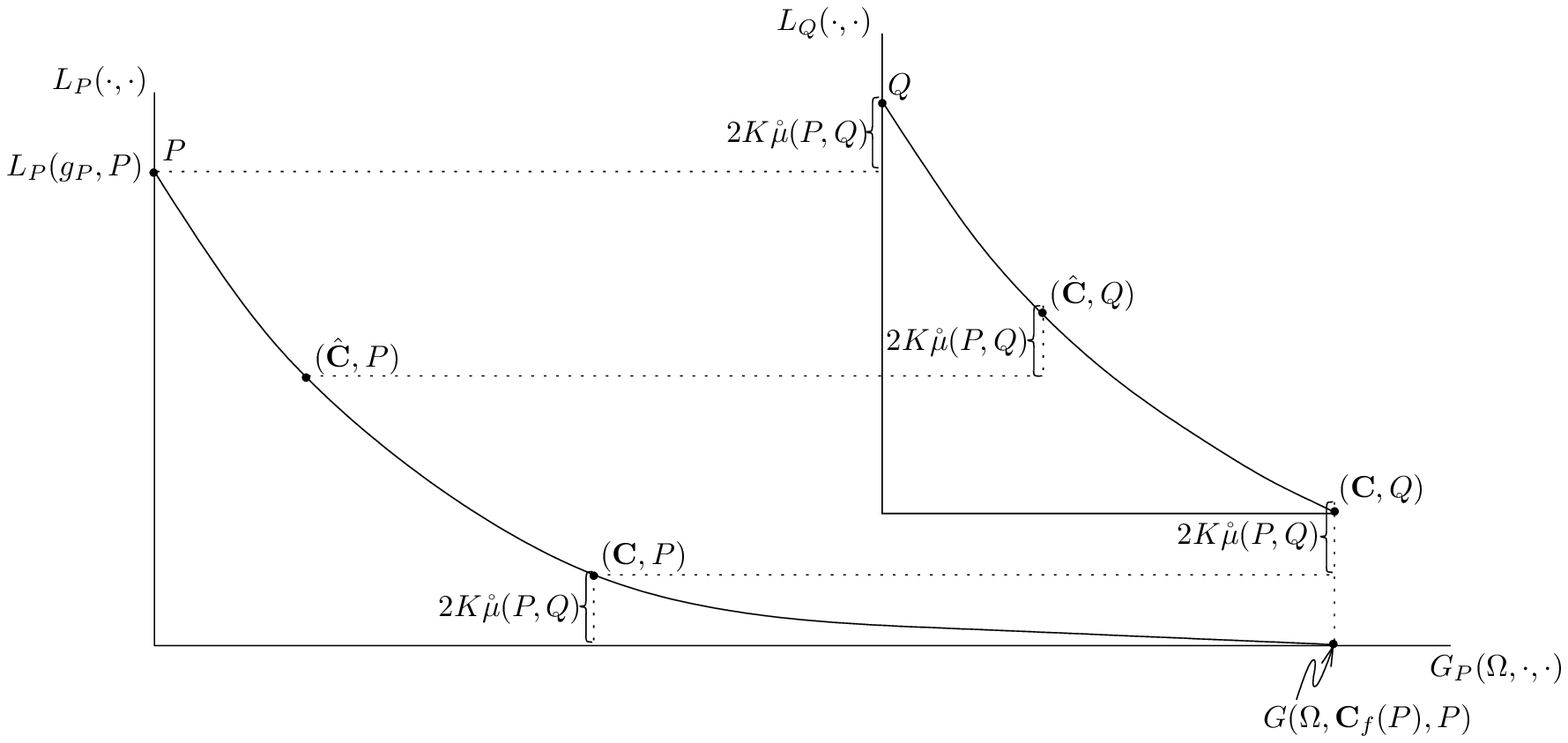}
} \caption{\label{f:lossub}Pseudoenergy (including virtual pseudoenergy) vs. relative loss.}
\end{figure}


Several useful observations can now be made.
\begin{itemize}
\item The result of Proposition~\ref{p:L<mu} suggests that good (near-optimal) partitions of $\Omega$ can be generated by a reduction of the original measure $P$ to a measure $Q$ that is (i) supported at a few points and (ii) has a low value of the Fortet-Mourier metric $\zeta_c(P,Q)=\hat\mu_{\hat c}(P,Q)$. The latter value of the Monge-Kantorovich functional $\hat\mu_{\hat c}(P,Q)$ with the reduced cost $\hat c$ can be computed as that of a minimum-cost transportation problem.

\item For a wide class of linear multi-period two stage stochastic optimization problems, the relevant cost function $c$ is given by $c_p$ (see~Appendix B) with $p=l+1$ where $l$ is the number of periods. The corresponding minimum cost transportation problem can easily be solved exactly for fixed support of measure $Q$ and approximately if the support itself needs to be optimized (see~Appendix C for details).

\item The optimality ``price'' one pays for scenario reduction from the original measure $P$ to a simpler measure $Q$ -- which can be thought of as adding information that's minimally relevant to the problem in question without actually finding it -- can be estimated by the amount $2K \hat \mu_{\hat c}(P,Q)$. This implies, in particular, that one could do a scenario reduction before starting the search for the efficient frontier. In fact, scenario reduction and additional information acquisition are complementary to each other in the sense of information: scenario reduction, as already mentioned, can be thought of as an addition of information that's minimally relevant as opposed to information acquisition optimization, where one looks for maximally relevant information.
\end{itemize}
It is now possible to formulate an efficient approximate algorithm for optimal partition determination.

\subsection{Efficient frontier algorithm}
Proposition~\ref{p:L<mu} provides a useful tool for approximating the efficient frontier.
Specifically, one can use the following algorithm (here and later we assume that the original measure $P$ on $\Omega$ has a support at a discrete set $\Omega_N\subset \Omega$ consisting of $N$ points).
\begin{enumerate}
\item Choose an integer parameter $\ge 2$.

\item Choose an appropriate cost function $c$: $\Omega\times\Omega\rightarrow \bR_+$ such that $f(\o,x)\in \mF_c$ for all $x\in X$. Let $\hat c$ be the corresponding reduced cost function.

\item  Reduce the original measure $P$ to measure $Q$ supported at $r$ points in the set $\Omega_N$, i.e. find a $\hat c$-optimal map $\nu\in \mR_r(\Omega_N)$ such that $Q=\nu(P)$.

\item Let $\bC$ be any partition of $\Omega$ generated by the map $\nu$.

\item Let the map $g_{\bC,P}\in \mC$ be a subset-optimal map corresponding to partition $\bC$.
\end{enumerate}

Varying the value of parameter $r$ from 2 upwards one can obtain a series of maps in the set $\mC$ that
are (approximately) Pareto-optimal. Step 2 of the above algorithm is essential for its feasibility. For example, if the problem (\ref{eq:gen_stoch}) is a linear multi-period stochastic optimization problem, the cost function of the form (\ref{eq:c_p}) can be used. In step 3, finding the measure $Q$ supported at $r$ points that minimizes the value of Monge-Kantorovich functional $\hat \mu_{\hat c}(P,Q)$ is an NP-hard problem \cite{heitsch2003} but approximate algorithm such as {\it fast forward selection algorithm} are available (see~Appendix C).

Using the algorithm described above, one can obtain one approximately Pareto-optimal map for each value of the chosen integer parameter. If more Pareto-optimal maps are needed (especially in the region with lower values of pseudoenergy) additional heuristics can be used. For instance, one could begin with the algorithm described above for some relatively high value of $r$ and then merge some of the resulting subsets into one giving rise to a partition with a lower value of $r$. Clearly, this can be done in $B_r-1$ ways, where $B_n$ is the $n$-th Bell number which is just the number of all different partitions of a set consisting of $n$ elements and that can be found from the recursive relation $B_{n+1}=\sum_{k=0}^n \binom{n}{k}B_{k}$ and $B_0=1$. (For example, the Bell number for the lower values of $n$ are $B_2=2$, $B_3=5$, $B_4=15$, $B_5=52$, $B_6=203$, $B_7=877$, $B_8=4140$.)

We see that if the original chosen value of $r$ is not very high this would lead to a manageable number of partitions. Additionally, scenario reduction can be used to reduce computational complexity of finding the values $x^*_{P_C}$ for subsets $C$ of resulting partitions. On the other hand, if the original value of $r$ makes evaluation of all maps that can be obtained this way computationally prohibitive, a heuristic algorithm described by the following pseudo-code can be used. It finds another partition, with a lower value of $r$, so that the subset merging procedure can be applied.


\begin{algorithm}
\caption{Approximation to Pareto-optimal boundary.}
\label{code1}
\SetKw{Input}{Input:}
\SetKw{Step}{Step $0$:}
\SetKw{Stepk}{Step $k = 1,\ldots,n$:}
\SetKw{Stepn}{Step $n+1$:}
\Input\;
\Indp
    ${\bf C} = \{C_1,\ldots,C_r\},$\;
    $\{\o_1,\ldots,\o_r|\o_i\in C_i\}\subset\Omega$,\;
    choose an integer $n$ such that $1\leq n \leq r-2$.\;
\Indm
\Step\;
\Indp
    $J^{[0]} := \{1,\ldots r\},$\;
    ${\bf C}' := \{C'_1,\ldots,C'_{n+1}\}$ such that $C'_i : = \emptyset,\,\forall\, i$,\;
    calculate $\hat{c}_p(\o_i,\o_j),\;\forall\, i,j\in J^{[0]}$.\;
\Indm
\Stepk\;
\Indp
        \ForEach{$i\in J^{[k-1]}$}{
            $\bar{c}_p(i) :=\frac{1}{|J^{[k-1]}|}\sum_{j\in J^{[k-1]}} \hat{c}_p(\o_i,\o_j)$,\;
            }
        $u_k := \operatorname*{arg\,max}_{i\in J^{[k-1]}}\bar{c}_p(i)$,\;
        $J^{[k]} := J^{[k-1]} \backslash \{u_k\}$,\;
        $C'_k:= C_{u_k}$.\;
\end{algorithm}

The goal of the algorithm represented by the pseudo-code is to identify subsets which are locally compact but as far away from one another as possible.
 In each step $k$, we find the average distance $\bar{c}_p$ of each
subset center remaining in the index set $J^{[k-1]}$ to only the other remaining centers.  The center, and
therefore the associated subset, with the largest average distance is chosen and removed from the set
$J^{[k-1]}$.  The remaining subsets are then merged into a single set.


So far the pseudotemperature function $u$ has not been taken into account. It is clear, on the other hand, that it will in general affect the composition of the set $\mO$ of Pareto-optimal maps. In order to properly incorporate the pseudotemperature function into the heuristics described above, one could note that the questions difficultly is generally smaller when subsets with high pseudotemperature values have large measures as well. In other words, if one wishes to keep the question difficulty low, one should avoid creating subsets of small measure in regions of the parameter space characterized with high pseudotemperature values. To facilitate creation of such subsets, one could, for example modify the (reduced) cost function $\hat c$ in the following way
\begin{equation}
\hat c(\o_i,\o_j)\rightarrow \frac{\hat c(\o_i,\o_j)}{f_c(u(\o_i),u(\o_j)},
\label{eq:mod-hatc}
\end{equation}
where $f_c$: $\bR_+\times\bR_+\rightarrow \bR_+$ is some increasing function of its arguments. The specific shape of $f_c$ can be determined experimentally, and several shapes can be tried for every given instance assuming computational resources are not a limiting factor.

\section{\label{s:example}Example}
Let us consider an example.
The original problem is a that of two-stage linear stochastic optimization with simple recourse taken from a well-known textbook \cite{BIRGE:1997}.  The problem is for a farmer to allocate the appropriate amount of land between wheat, corn and sugar beets in order to maximize profits.  The farmer knows that at least 200 tons of wheat and 240 tons of corn must be grown for cattle feed.  If not enough is grown to satisfy this demand, both wheat and corn can be bought for \$238 and \$210 per ton, respectively.  Any excess above the demand can be sold for \$170 and \$150 per ton of wheat and corn, respectively.  It costs \$150 per acre to plant the wheat and \$230 per acre to plant the corn.  The farmer can also grow sugar beets that sell for \$36 per ton. However, there is a quota of 6000 tons and any amount grown above this may only be sold at \$10 per ton.  It costs \$260 per acre to plant sugar beets.  The farmer has 500 acres available.

The problem can be stated as:
\begin{align*}
\textrm{minimize}\quad   & 150x_1 + 230x_2 + 260x_3 + \mathbb{E}_P Q(x,\Omega) \label{FP}\tag{FP}\\
\textrm{subject to}\quad & x_1+x_2+x_3 \quad \le 500\\
                         & x_1,x_2,x_3 \ge 0,
\end{align*}
where the second stage problem for a specific scenario can be written
\begin{align*}
Q(x,s) = \textrm{minimize}&\{238y_1 - 170w_1 + 210y_2 - 150w_2 - 36w_3 - 10w_4\}\\
\textrm{subject to} \quad & \omega_1(s)x_1 + y_1 + w_1 \quad \ge 200\\
                          & \omega_2(s)x_2 + y_2 + w_2   \quad \ge 240\\
                          & w_3 + w_4          \quad \le \omega_3(s)x_3\\
                          & w_3                \quad \le 6000\\
                          & y_1,y_2,w_1,w_2,w_3,w_4 \ge 0
\end{align*}
where $\omega_i(s)$ represents the yield of crop $i := 1,2,3$ for wheat, corn, and sugar beets, respectively, under scenario $s$; $x_i$ are the acres of land to devote to each crop $i$; $y_1, y_2$, are tons of wheat and corn, respectively, purchased to meet cattle feed requirements; $w_1, w_2, w_3, w_4$ are tons of wheat, corn, sugar beets below quota, and sugar beets above quota, respectively, sold for profit.

The problem has been modified in order to create the illustrative example used below.  In this example, only wheat and sugar beet yields are uncertain.  Each is allowed to take five different values of yields resulting in 25 scenarios.  For the sake of convenience, we assume that the corn yield is non-random and is equal to 3 tons per acre, while for both wheat and beets the average yield equal to 2.5 and 20, respectively, has a probability of 0.30. The yield for both of these cultures can be either higher or lower than average by 20\% with probability 0.20 and also can be higher or lower than average by 30\% with probability 0.15. The yields for wheat and beets are assumed to be independent.

The resulting uncertain yields are summarized below:

\begin{tabular}{lll}
wheat       & [1.75, 2.00, 2.50, 3.00, 3.25] & w.p. (0.15,0.20,0.30,0.20,0.15)\\
corn        & [3] & w.p. (1)\\
beets  & [14, 16, 20, 24, 26]& w.p. (0.15,0.20,0.30,0.20,0.15)\\
\end{tabular}

\begin{figure}
\centerline{
\includegraphics*[scale=0.48]{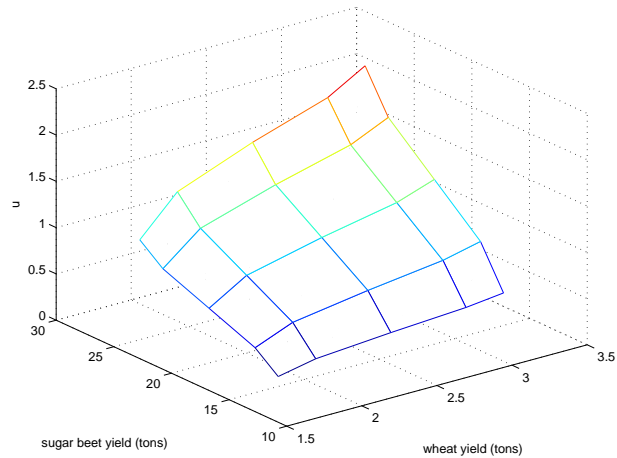}
}
\caption{\label{f:ptemp}Pseudotemperature function given for the farmer land allocation problem with uncertainty residing in the yields of wheat and sugar beets.}
\end{figure}

Also, let us assume that the pseudotemperature function $u(\o_1,\o_2)$ is given as
\begin{align}
u(i,j) = i\cdot j^{0.5}, \forall\quad i,j \in {1,\ldots,5}
\end{align}
where $i,\;j$ are the indices referencing the uncertain yields of wheat and sugar beets, respectively (where the smallest value of the uncertain yield corresponds to $i=1$ ($j=1$) and the largest yield corresponds to $i=5$ ($j=5$)).  The pseudotemperature function is then normalized so that $\bE_P u(ij) = 1$.  Fig.~\ref{f:ptemp} shows a plot of the pseudotemperature function.

\begin{figure}
\centerline{
\includegraphics*[scale=0.50]{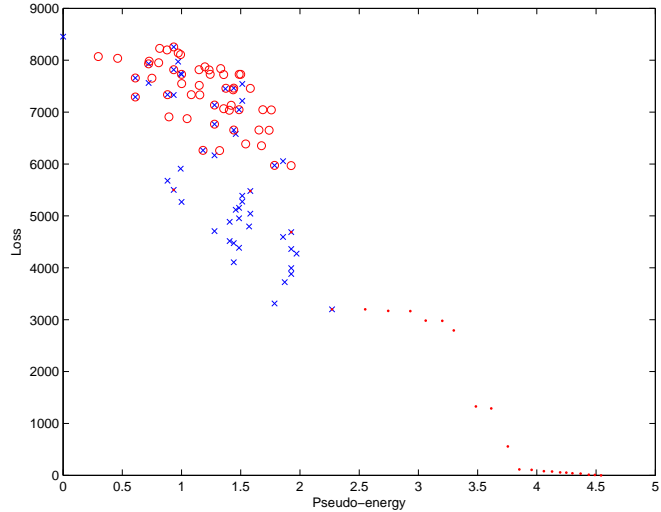}
}
\caption{\label{f:scatter}Maps that are generated by scenario reduction for various values of $r$ (solid dots), scenario reduction for $r=5$ with subsequent subset merging (crosses), scenario reduction to $r=10$, reducing to $r=5$ using the pseudo-code and subsequent subset merging (circles). Pseudotemperature function is set to a constant.}
\end{figure}

\begin{figure}
\centerline{
\includegraphics*[scale=0.50]{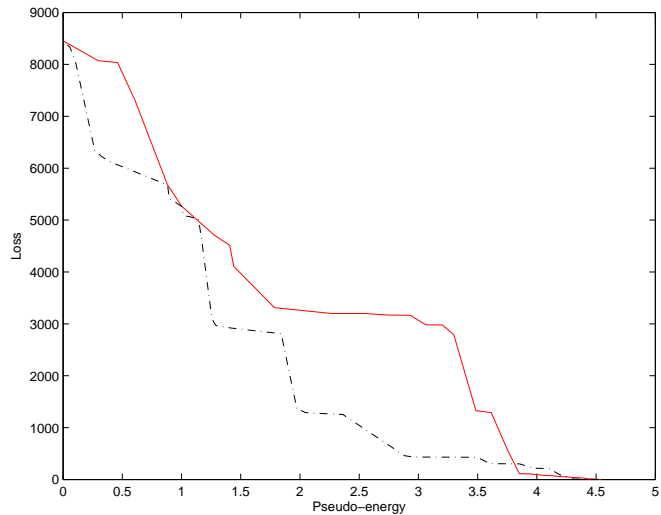}
}
\caption{\label{f:eff-front-2}Approximate efficient frontiers for the constant pseudotemperature function (solid line) and pseudotemperature shown in Fig.~\ref{f:ptemp}.}
\end{figure}

The efficient frontier can be approximated by using the scenario reduction based algorithm described in the previous section together with subset merging heuristics. The resulting maps are shown in Fig.~\ref{f:scatter} for the case of constant pseudotemperature. The resulting approximate efficient frontier both for constant pseudotemperature function and for the pseudotemperature given shown in Fig.~\ref{f:ptemp} are shown in Fig.~\ref{f:eff-front-2}.

Now consider an information source described by the modified linear model with parameters $b=0.8$ and $Y_s=0.2$ (which is a rather modest capacity value). We would like to find out how much the original loss can be reduced by optimally using such an information source. In other words, we want to solve problem (\ref{eq:minLV}). For this purpose one can take questions on the (approximate) efficient frontier and plot parametric curves\\ $(Y(\Omega,\bC,P,V_{\alpha}(\bC)), \mL(V_{\alpha}(\bC)))$ where $\mL(V_{\alpha}(\bC))$ is given by Proposition~\ref{p:loss-imp}. The question yielding the lowest point of intersection of such a curve with the vertical line $G=Y_s$ will give an approximate solution of problem (\ref{eq:minLV}).

\begin{figure}
\centerline{
\includegraphics*[scale=0.50]{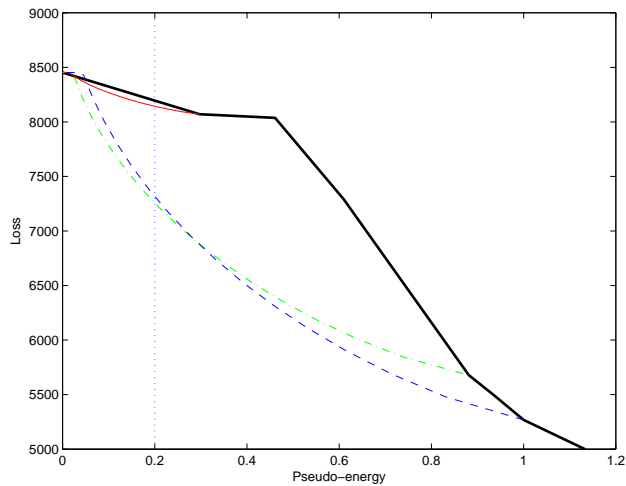}
}
\caption{\label{f:res-const-u}Part of approximate efficient frontier and parametric loss curves for quasi-perfect answers to three different questions for the case of constant pseudotemperature.}
\end{figure}

Results for the case of constant pseudotemperature are shown in Fig.~\ref{f:res-const-u}. The parametric curves for three questions (all three with $r=2$) are produced. We can see that the lowest value of the expected loss that can be obtained this way is equal to 7250 which constitutes a reduction of about 14\%.

\begin{figure}
\centerline{
\includegraphics*[scale=0.50]{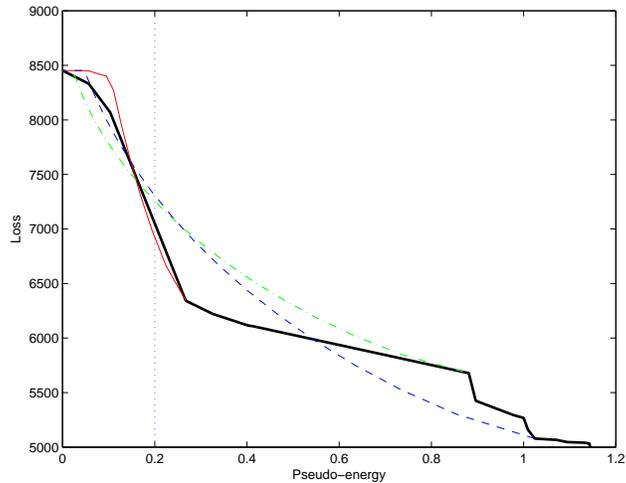}
}
\caption{\label{f:res-var-u}Part of approximate efficient frontier and parametric loss curves for quasi-perfect answers to three different questions for the case of non-constant pseudotemperature shown in Fig.~\ref{f:ptemp}.}
\end{figure}

For the case of non-constant pseudotemperature are shown in Fig.~\ref{f:res-var-u}. Analogously, three $r=2$ questions were chosen on the approximate efficient frontier and the corresponding parametric curves plotted. The best curve is observed to intersect the vertical line $G=0.2$ at the value of vertical coordinate equal to about 6900 which represents a reduction of about 18\% compared to the EVPI of 8450 of the original problem.

\section{\label{s:conclusion}Conclusion}
The main subject of this work is the  development of approximate methods for solving the problem of optimizing additional information acquisition in decision making problems with uncertainty that are typically solved using stochastic optimization techniques. It represents a logical continuation of the developments presented in \cite{loss-part1}. The main problem that was formulated \cite{loss-part1} is that of finding the efficient frontier in the space of questions to the given information source and determining optimal question(s) that allow for the loss reduction maximization for the problem the agent is interested in solving.

The solution methods proposed in here are based on the method of probability metrics and their application for scenario reduction in stochastic optimization (see appendices). The main idea is that, informally speaking, optimal scenario reduction on one hand and optimal information acquisition on the other hand are complementary processes. More specifically, in scenario reduction, the goal is to reproduce the overall shape of the original probability distribution as faithfully as possible with a probability measure of a smaller support -- one strives to keep the ``overall shape'' of the distribution while leaving out the ``small details''. From the informational point of view, this corresponds to searching for the {\it least relevant} information and adding it (updating the measure accordingly) -- without finding it.
In information acquisition, on the contrary, the goal is to find the {\it most relevant} information that, at the same time, is relatively easy (so that the question requesting it has sufficiently low difficulty)
for the information source to supply -- and, hence, it ends up being accurate. Therefore, if one has a method for finding the least relevant information, the same method can likely be made to work for finding the most relevant information.

This observation provides for a means of development of simple approximate algorithms for determining the efficient frontier and for finding optimal questions for the given information source. The methods described here work for the class of linear multi-period two stage stochastic optimization problems and should generalize relatively easily to other problem classes for which scenario reduction based on probability metrics was shown to be possible, including chance constrained and two-stage integer stochastic optimization problems.

\appendix

\section{\label{a:proofs}Proofs}

\subsection{Proof of Lemma~\ref{l:av<2min}}
Let $i^*=\arg\min_{i}\sum_{j=1}^N p_jc_{ij}$ (so that $\min_{i} \sum_{j=1}^N p_jc_{ij}=\sum_{j=1}^N p_j c_{i^*j}$). Then we can write
\begin{align*}
\sum_{i=1}^N \sum_{j=1}^N p_ip_j c_{ij} &\overset{(a)}{\le} \sum_{i=1}^N \sum_{j=1}^N p_ip_j (c_{ii^*}+c_{i^*j})\\
 &= \sum_{i=1}^N\sum_{j=1}^N p_ip_j c_{ii^*} + \sum_{i=1}^N\sum_{j=1}^N p_ip_j c_{i^*j}\\
 &= \sum_{j=1}^N p_j \sum_{i=1}^N p_i c_{i^*i} + \sum_{i=1}^N p_i \sum_{j=1}^N p_jc_{i^*j} \\
 &\overset{(b)}{=} 2\min_i \sum_{j=1}^N p_jc_{ij},
\end{align*}
where (a) follows from the triangle inequality satisfied by the elements $c_{ij}$ and (b) follows from the definition of $i^*$.

\subsection{Proof of Lemma~\ref{l:zeta_c}:}
The first statement actually holds true for any measures $P, Q\in \mP_c(\Omega)$ (see Appendix~1 for the definition of $\mP_c(\Omega)$). Indeed, let $f^*(\o)\in \mF_c$ be the function that achieves the maximum of
$$\left\vert \int_{\Omega} f(\o) P(d\o)-\int_{\Omega}
 f(\o) Q(d\o) \right\vert . $$
Let $f^*_j(\o)$ be the restriction of $f^*(\o)$ to $C_j$. Clearly, $f^*_j(\o)\in \mF_c(C_j)$. We can write
\begin{align*}
\zeta_c(P,Q) &= \left\vert \int_{\Omega} f^*(\o) P(d\o)-\int_{\Omega}
 f(\o) Q(d\o) \right\vert\\
&\overset{(a)}{=} \sum_{j=1}^r w_j \left\vert \int_{C_j} f^*(\o) dP_{C_j}(\o) - \int_{C_j} f^*(\o) dQ_{C_j}(\o) \right\vert \\
&\overset{(b)}{=} \sum_{j=1}^r w_j \left\vert \int_{C_j} f^*_j(\o) dP_{C_j}(\o) - \int_{C_j} f^*_j(\o) dQ_{C_j}(\o) \right\vert\\
 &\overset{(c)}{\le} \sum_{j=1}^r w_j \zeta_c(P_{C_j}, Q_{C_j}),
\end{align*}
where (a) follows from the definition of conditional measures $P_{C_j}$ and $Q_{C_j}$, (b) follows from the definition of functions $f^*_j(\o)$, and (c) follows from that $f^*_j(\o)\in \mF_c(C_j)$ and definition of $\zeta_c(P_{C_j}, Q_{C_j})$.

To prove the second statement, we can use the duality result (\ref{eq:KR}) together with (\ref{eq:KR=MK}) that relates the values of Kantorovich-Rubinstein and Monge-Kantorovich functionals. Let $\nu\in \mR_r(\Omega_N)$ be the map that generates partition $\bC$, and let $\tilde\o_j=\nu(\o_i)$ for all $\o_i\in \bC_j$. Note also that $q_j=\sum_{\{i:\o_i\in C_j\}}p_i=w_j$, $j=1,\dotsc, r$. We can write
\begin{align*}
\sum_{j=1}^r w_j\zeta_c(P_{C_j},Q_{C_j})&\overset{(a)}{=}\sum_{j=1}^r w_j\hat\mu_{\hat c}(P_{C_j},Q_{C_j})
\overset{(b)}{=} \sum_{j=1}^r w_j \sum_{\{i:\o_i\in C_j\}}\frac{p_i}{w_j}\hat c(\o_i,\tilde\o_j)\\
&= \sum_{j=1}^r \sum_{\{i:\o_i\in C_j\}} p_i\hat c(\o_i, \tilde\o_j) \overset{(c)}{=} \hat\mu_{\hat c}(P,Q)\overset{(d)}{=}\zeta_c(P,Q),
\end{align*}
where (a) and (d) follow from (\ref{eq:KR}) and (\ref{eq:KR=MK}), (b) follows from that $Q_{C_j}$ is supported at a single point $\tilde\o_j$, (c) follows from the way measure $Q$ was constructed as a reduction of the measure $P$ with a $\hat c$-optimal map $\nu\in \mR_r(\Omega_N)$.

\subsection{Proof of Proposition~\ref{p:L<L+mu}:} Let $w_j=P(\hat C_j)=Q(\hat C_j)$ be the measure of subsets in $\hat\bC$ and let $P_j\equiv P_{\hat C_j}$ and $Q_j\equiv Q_{\hat C_j}$ be the corresponding subset measures.
\begin{align*}
 & L_P(\hat\bC,Q) = \sum_{j=1}^r w_j \left[\int_{\hat C_j} f(\o, x_{Q_j}^*) P_j(d\o) - \int_{\hat C_j} f(\o, x_{\o}^*) P_j(d\o)\right]\\
   &=\sum_{j=1}^r w_j \left[ \int_{\hat C_j} f(\o, x_{Q_j}^*) P_j(d\o) - \int_{\hat C_j} f(\o, x_{\o}^*) P_j(d\o)\right.\\
       &\left. + \int_{\hat C_j} f(\o, x_{P_j}^*) P_j(d\o) - \int_{\hat C_j} f(\o, x_{P_j}^*) P_j(d\o)\right]\\
       &\overset{(a)}{=} L_P(\hat\bC ,P) + \sum_{j=1}^r w_j \left[ \int_{\hat C_j} f(\o, x_{Q_j}^*) P_j(d\o) - \int_{\hat C_j} f(\o, x_{P_j}^*) P_j(d\o) \right] \\
       &= L_P(\hat\bC,P) + \sum_{j=1}^r w_j \left[ \int_{\hat C_j} f(\o, x_{Q_j}^*) P_j(d\o) - \int_{\hat C_j} f(\o, x_{P_j}^*) P_j(d\o) \right. \\
      & \left. \hspace{25mm} + \int_{\hat C_j} f(\o, x_{Q_j}^*) Q_j(d\o) - \int_{\hat C_j} f(\o, x_{Q_j}^*) Q_j(d\o) \right] \\
      &\overset{(b)}{=} L_P(\hat \bC,P) +\sum_{j=1}^r w_j \left[ v(Q_j)-v(P_j) + \int_{\hat C_j} f(\o, x_{Q_j}^*) (P_j-Q_j) (d\o) \right] \\
      &\le L_P(\hat\bC,P) + \sum_{j=1}^r w_j \left|v(Q_j)-v(P_j) \right| + \sum_{j=1}^r w_j \left| \int_{\hat C_j} f(\o, x_{Q_j}^*) (P_j-Q_j) (d\o) \right|\\
      &\overset{(c)}{\le} L_P(\hat \bC,P) + K\sum_{j=1}^r w_j\zeta_c(P_j,Q_j)+ K\sum_{j=1}^r w_j\zeta_c(P_j,Q_j)\\
      &= L_P(\hat \bC,P) + 2K\sum_{j=1}^r w_j \zeta_c(P_j,Q_j)\\
&\overset{(d)}{=} L_P(\hat \bC,P) +2K\zeta_c(P,Q)
\end{align*}
where (a) follows from the definition of $L_P(\hat\bC,P)$, (b) follows from the definition of the optimal objective values $v(P_j)$ and $v(Q_j)$, (c) follows from that the integrand $f(\o,x)$ is in $\mF_c$ and definition (\ref{eq:zeta_c}) of Fortet-Mourier metric $\zeta_c$, and (d) follows from Lemma~\ref{l:zeta_c}.

\subsection{Proof of Proposition~\ref{p:L<mu}:}
Let $w_j=P(C_j)=P(Q_j)$, $j=1,\dotsc, r$  be measures of subsets in $\bC$ and let $P_j$ and $Q_j$ be the corresponding subset measures.
\begin{align*}
&L(g_{\bC,P},P) =  \sum_{j=1}^r w_j L( g_{P_j},P_j)= \sum_{j=1}^rw_j \int_{C_j} \left(f(\o,x_{P_j}^*) - f(\o,x_{\o}^*)\right) P_j(d\o)\\
  &=\sum_{j=1}^r w_j \sum_{\{i:\o_i\in C_j\}} \frac{p_i}{w_j}\left( f(\o_i,x_{P_j}^*) - f(\o_i,x_{\o_i}^*)\right)\\
&\overset{(a)}{=} \sum_{j=1}^r w_j \left( v(P_j) - \sum_{\{i:\o_i\in C_j\}} (P_j)_i\, v(\delta_{\o_i}) \right)\\
&= \sum_{j=1}^r w_j \sum_{\{i:\o_i\in C_j\}} (P_j)_i  \left( v(P_j) - v(\delta_{\o_i}) \right)\\
&\overset{(b)}{\le} K\sum_{j=1}^r w_j \sum_{\{i:\o_i\in C_j\}} (P_j)_i\, \zeta_c(P_j,\delta_{\o_i})
\overset{(c)}{=} K\sum_{j=1}^r w_j \sum_{\{i:\o_i\in C_j\}} (P_j)_i\, \hat\mu_{\hat c}(P_j,\delta_{\o_i})\\
&=K\sum_{j=1}^r w_j \sum_{\{i:\o_i\in C_j\}} (P_j)_i\, \sum_{\{k:\o_k\in C_j\}} (P_j)_k \,
\hat c(\o_i,\o_k)\\
&\overset{(d)}{\le}  2K\sum_{j=1}^r w_j \min_{\{k:\o_k\in C_j\}} \sum_{\{i:\o_i\in C_j\}} (P_j)_i\, \hat c(\o_i,\o_k)
= 2K\sum_{j=1}^r w_j \min_{\{k:\o_k\in C_j\}} \hat\mu_{\hat c}(P_j, \delta_{\o_k})\\
&\overset{(e)}{=}2K\sum_{j=1}^r w_j \hat\mu_{\hat c}(P_j, Q_j)
=2K\sum_{j=1}^r w_j \zeta_c(P_j, Q_j) \overset{(f)}{=} 2K \zeta_c(P,Q),
\end{align*}
where $(P_j)_i\equiv \frac{p_i}{w_j}$ for $\o_i\in C_j$, (a) follows from the definition of optimal values $v(P_j)$ and $v(\delta_{\o_i})$, (b) follows from the upper bound (\ref{eq:stability}), (c) follows from the duality relation (\ref{eq:KR}) and from the relation (\ref{eq:KR=MK}) between the Kantorovich-Rubinstein and Monge-Kantorovich functionals, (d) follows from Lemma \ref{l:av<2min} (since $\hat c$ is a metric and $\{(P_j)_i\}_{\{i:\o_i\in C_j\}}$ is a probability distribution), (e) follows from that $Q=\nu(P)$, where $\nu$ is $\hat c$-optimal, and (f) follows from Lemma~\ref{l:zeta_c}.\qed

\section{\label{a:metrics}Probability metrics and stability in stochastic optimization}
Consider the problem  (\ref{eq:gen_stoch}). Let $\mP(\Omega)$ be the set of all Borel probability measures on $\Omega$ and define
\begin{equation*}
v(P)=\inf\left\{\int_{\Omega} f(\o,x)\,dP(\o)\, :\, x\in X \right\}
\end{equation*}
and
\begin{equation*}
S(P)=\left\{x\in X\, :\, \int_{\Omega} f(\o,x)\,dP(\o)=v(P) \right\}
\end{equation*}
to be the optimal value and optimal solution set of (\ref{eq:gen_stoch}), respectively.

Let's  also define (as in, for example, \cite{romisch2007})
\begin{equation*}
\mF=\left\{f(\cdot,x):\; x\in X \right\}
\end{equation*}
and
\begin{align*}
\mP_{\mF}(\Omega) &= \left\{ Q\in \mP:\; -\infty < \int_{\Omega}\inf_{x\in X \cap \rho\bB} f(\o,x)Q(d\o)\;\; \text{and} \right. \\
&  \left. \sup_{x\in X\cap \rho\bB } \int_{\Omega} f(\o,x)Q(d\o)<\infty,\; \text{for all}\, \rho>0 \right\},
\end{align*}
where $\bB$ is the closed unit ball in $\bR^n$.

Then the probability distance of the form
\begin{equation}
d_{\mF,\rho}(P,Q)=\sup_{x\in X\cap \rho\bB} \left\vert \int_{\Omega} f(\o,x) P(d\o)-\int_{\Omega}
 f(\o,x) Q(d\o) \right\vert
 \label{eq:pr_metric}
 \end{equation}
can be defined on $\mP_{\mF}(\Omega)$. This distance is called  {\it Zolotarev's pseudometric with $\zeta$-structure} \cite{zolotarev1983,rachev2002,rachev1998a,rachev1998b}. The pseudometric (\ref{eq:pr_metric}) would become a metric if the class $\mF$ were rich enough so that $d_{\mF,\rho}(P,Q)=0$ implies $P=Q$.

Theorem 2 in \cite{dupacova2003} states that if $P,Q\in \mP_{\mF}$, $S(P)$ is nonempty and bounded then there exist $\rho>0$ and $\delta>0$ such that
\begin{equation}
|v(P)-v(Q)|\le d_{\mF,\rho}(P,Q)
\label{eq:ub}
\end{equation}
is valid for all $Q\in \mP_{\mF}$ such that $d_{\mF,\rho}(P,Q)<\delta$.

The distance $d_{\mF,\rho}$ in (\ref{eq:ub}) is typically difficult to handle since the class of functions $\mF$  is determined by the specific integrand $f(\o,x)$ for the given instance of problem (\ref{eq:gen_stoch}). The main idea underlying the use of the probability metrics method for the study of stability and for scenario reduction in stochastic programming is to suitably enlarge the class $\mF$ so that it still shares its main analytical properties with functions $f(\cdot, x)$. Such properly enlarged classes are sometimes referred to as {\it canonical classes} and the corresponding metrics are sometimes called {\it canonical metrics}.

Consider, for instance the class $\mF_c$ of continuous functions defined as
\begin{equation}
\mF_c = \left\{f:\; \Omega\rightarrow \bR :\; |f(\o)-f(\tilde\o)|\le c(\o,\tilde\o),\; \text{for all}\; \o,\tilde\o \in \Omega \right\},
\label{eq:F_c}
\end{equation}
where $c$: $\Omega\times\Omega\rightarrow \bR_+$ is a continuous symmetric function such that $c(\o,\tilde\o)=0$ if and only if $\o=\tilde\o$.
Then the corresponding (pseudo-) metric has the form
\begin{equation}
\zeta_c(P,Q)\equiv d_{\mF_c}(P,Q)=\sup_{f\in \mF_c} \left\vert \int_{\Omega} f(\o) P(d\o)-\int_{\Omega}
 f(\o) Q(d\o) \right\vert
 \label{eq:zeta_c}
 \end{equation}
and is known as {\it Fortet-Mourier metric}. If the cost function $c(\o,\tilde\o)$ satisfies additional boundedness and continuity conditions:
\begin{itemize}
\item $c(\o,\tilde\o)\le \lambda(\o)+\lambda(\tilde\o)$ for some $\lambda$: $\Omega\rightarrow \bR_+$ mapping bounded sets into bounded sets,
\item $\sup\{c(\o,\tilde\o)\; :\; \o,\tilde\o\in \bB_{\epsilon}(\o_0), ||\o,-\tilde\o||\le \delta \}\rightarrow 0$ as $\delta\rightarrow 0$ for each $\o_0\in \Omega$, where $\bB_{\epsilon}(\o_0)$ is the $\epsilon$-ball centered at $\o_0$,
\end{itemize}
the Fortet-Mourier metric (\ref{eq:zeta_c}) admits a dual representation as
the {\it Kantorovich-Rubinstein functional} \cite{rachev1990}:
\begin{equation}
\begin{split}
\zeta_c(P,Q)=\overset{\circ}{\mu}_c(P,Q)&=\inf\left\{\int_{\Omega\times\Omega}c(\o,\tilde\o)\eta(d\o, d\tilde\o) : \eta\in \mP(\Omega\times \Omega),\right.\\
&\left.\pi_1\eta-\pi_2\eta=P-Q \right\},
\end{split}
\label{eq:KR}
\end{equation}
where $\pi_1$ and $\pi_2$ denote projections on first and second components, respectively.
It is straightforward to show that the Kantorovich-Rubinstein functional (\ref{eq:KR}) can be upper-bounded by the {\it Monge-Kantorovich functional}:
\begin{equation}
\begin{split}
\overset{\circ}{\mu}_c(P,Q)\le \hat\mu_c(P,Q)&=\inf\left\{\int_{\Omega\times\Omega}c(\o,\tilde\o)\eta(d\o, d\tilde\o) : \eta\in \mP(\Omega\times \Omega),\right.\\
&\left.\pi_1\eta=P, \pi_2\eta=Q \right\},
\end{split}
\label{eq:MK}
\end{equation}
and that the bounds becomes tight, (i.e.
$\overset{\circ}{\mu}_c(P,Q)= \hat\mu_c(P,Q)$)
if the cost function $c(\o,\tilde\o)$ is a metric on $\Omega$ \cite{levin1975}.
The problem of finding the minimum in (\ref{eq:MK}) is known the {\it Monge-Kantorovich mass transportation problem}.

Note that if measures $P$ and $Q$ are discrete
 ($P=\sum_{i=1}^N p_i\delta_{\o_i}$ and $Q=\sum_{j=1}^M q_i\delta_{\tilde\o_j}$), the Monge-Kantorovich functional (\ref{eq:MK}) takes the following form:
\begin{equation}
 \begin{split}
 \hat\mu_c(P,Q) &= \min \left\{\sum_{i=1}^N \sum_{j=1}^M c(\o_i, \tilde\o_j)\eta_{ij} : \eta_{ij}\ge 0,
 \sum_{i=1}^N \eta_{ij} =q_j, \sum_{j=1}^M \eta_{ij}=p_i\; \forall i,j\right\} \\
           &= \max\left\{\sum_{i=1}^N p_iu_i+\sum_{j=1}^M q_jv_j : u_i+v_j\le c(\o_i,\tilde\o_j)\;
           \forall i,j \right\}
 \end{split}
 \label{eq:MKd}
\end{equation}

Given the cost function $c(\o,\tilde\o)$ one can define the {\it reduced cost} $\hat c(\o, \tilde\o)$ on $\Omega\times \Omega$ by
\begin{equation}
\hat c(\o, \tilde\o)=\inf \left\{\sum_{i=1}^{m-1} c(\o_i, \o_{i+1}) : m\in \mathbb{N},\, \o_i\in \Omega,\, \o_1=\o,\, \o_m=\tilde\o \right\}.
\label{eq:reduced-c}
\end{equation}
In can easily be shown that the reduced cost function $\hat c(\o,\tilde\o)$ is a metric (since it satisfies the triangle inequality) on $\Omega$ and that $\hat c(\o,\tilde\o)\le c(\o,\tilde\o)$ with the inequality being tight when $c(\o,\tilde\o)$ is also a metric.

It can also be shown (see \cite{rachev1998a}, chapter 4) that if $\Omega$ is compact with analytic sublevel sets then the Kantorovich-Rubinstein functional (\ref{eq:KR}) with the reduced cost function $\hat c$ coincides with the Kantorovich-Rubinstein functional with the original cost function $c$ (the result referred to as the {\it reduction theorem}):
\begin{equation}
\overset{\circ}{\mu}_{\hat c}(P,Q)=\overset{\circ}{\mu}_c(P,Q).
\label{eq:reduction-t}
\end{equation}
Since the reduced cost is a metric on $\Omega$ we have $\overset{\circ}{\mu}_{\hat c}(P,Q)= \hat\mu_{\hat c}(P,Q)$ and, comparing with (\ref{eq:reduction-t}) we conclude that, for compact parameter spaces with analytic sublevel sets, the equality
\begin{equation}
\overset{\circ}{\mu}_{c}(P,Q)=\hat\mu_{\hat c}(P,Q)\le \hat\mu_{c}(P,Q)
\label{eq:KR=MK}
\end{equation}
holds true.

We thus arrive at the following useful stability result. If the integrand in problem (\ref{eq:gen_stoch}) belongs to class $\mF_c$ for all $x\in X$ for some cost function $c$ satisfying additional boundedness and continuity conditions described earlier in the appendix, then the estimate
\begin{equation}
|v(P)-v(Q)|\le \zeta_c(P,Q)=\overset{\circ}{\mu}_{c}(P,Q)=\hat\mu_{\hat c}(P,Q)
\label{eq:stability}
\end{equation}
is valid for Borel measures $P$ and $Q$ in $\mP_c(\Omega)$ on compact $\Omega$ characterized with analytic sublevel sets. (Here $\mP_c(\Omega)=\{Q\in \mP(\Omega)\, :\, \int_{\Omega} c(\o,\o_0)dQ(\o)<\infty \}$ for some $\o_0\in \Omega$.)

The particular function $c(\o,\tilde\o)$ that plays an important role in the context of convex stochastic optimization has the form
\begin{equation}
c_p(\o, \tilde\o)=\max\{1, ||\o-\o_0 ||^{p-1}, ||\tilde\o-\o_0 ||^{p-1}\}||\o-\tilde\o ||,
\label{eq:c_p}
\end{equation}
for some $\o_0\in \Omega$.
The corresponding metric $\zeta_p\equiv \zeta_{c_p}$ is referred to as the {\it $p$-th order Fortet-Mourier metric}.

To give an example of a class of problems for which the $p$-th order Fortet-Mourier metric is relevant, consider linear multi-period stochastic optimization problems of the form
\begin{equation}
\begin{split}
 &\min \left\{ cy_0+\bE_P\Bigl(\min \sum_{j=1}^l c_j(\o)y_j\Bigr) : y_0\in X,\, y_j\in Y_j,\, \right.\\ &\left.W_{jj}y_j=b_j(\o)-W_{jj-1}(\o)y_{j-1},\, j=1,\dotsc,l \right\},
\end{split}
 \label{eq:multiperiod}
\end{equation}
where $Y_j\subseteq \bR^{n_j}$ are polyhedral sets. Problem (\ref{eq:multiperiod}) can be written in the form (\ref{eq:gen_stoch}) with the integrand $f(\o,x)$ given by
\begin{align*}
f(\o,x) &= cx+ \inf \left\{ \sum_{j=1}^l c_j(\o)y_j : y_j\in Y_j,\, W_{jj}y_j=b_j(\o)-W_{jj-1}(\o)y_{j-1},\,\right.\\
&\left. j=1,\dotsc,l  \right\} = cx+\Psi_1(\o,x),
\end{align*}
where the function $\Psi_1(\o,x)$ is defined recursively:
\begin{equation*}
\Phi_j(\o,u_{j-1})=\inf \left\{c_j(\o)y_j+\Psi_{j+1}(\o,y_j) : y_j\in Y_j,\, W_{jj}y_j=u_{j-1} \right\}
\end{equation*}
\begin{equation*}
\Psi_j(\o, y_{j-1})=\Phi_j(\o, b_j(\o)-W_{jj-1}(\o)y_{j-1})
\end{equation*}
for $j=l,\dotsc,1$ and $\Psi_{l+1}(\o,y_l)\equiv 0$.

It is shown in \cite{romisch2007} that if $b_j(\o)-W_{jj-1}(\o)x\in W_{jj}Y_j$ for all pairs $(\o, x)$ ({\it relatively complete recourse}) and $\ker(W_{jj})\cap Y_j^{\infty}=\{0\}$ for $j=1,\dotsc, l-1$ (where $Y_j^{\infty}$ denotes the horizon cone\footnote{The horizon cone $D^{\infty}$ for the convex set $D\subseteq \bR^m$ is defined as the set of all elements $x_d\in \bR^m$ such that $x+\lambda x_d\in D$ for all $x\in D$ and all $\lambda\in \bR_+$. In particular, $D^{\infty}=\{0\}$ if $D$ is bounded.} of $Y_j$) then there exists a constant $\hat K$ such that
\begin{equation}
|f(\o,x)-f(\tilde\o,x)|\le \hat K \max\{1,\rho, ||\o||^l, ||\tilde\o||^l \}||\o-\tilde\o||
\label{eq:Lipsch}
\end{equation}
for all $\o$, $\tilde\o\in \Omega$ and $x\in X\cap \rho\bB$. This implies that $\frac{1}{\hat K\max\{1,\rho\}}f(\o,x)\in \mF_{c_{l+1}}$ for all $\o$, $\tilde\o\in \Omega$ and $x\in X\cap \rho\bB$.

It is now straightforward to obtain the following result (\cite{romisch2007}). Let $v(P)$ be the optimal value of problem (\ref{eq:multiperiod}). Assume that the relatively complete recourse condition for (\ref{eq:multiperiod}) is satisfied and that $\ker(W_{jj})\cap Y_j^{\infty}=\{0\}$ for $j=1,\dotsc, l-1$. Then there exists a constant $K>0$ such that the estimate
\begin{equation}
|v(P)-v(Q)|\le K \zeta_{l+1}(P,Q)
\label{eq:ubzeta}
\end{equation}
is valid for any $P$, $Q\in \mP_{l+1}(\Omega)$. (Here $\mP_{l+1}(\Omega)$ denotes the set of Borel measures on $\Omega$ with finite $(l+1)$-th order moments.)

Specifying the general result (\ref{eq:stability}) to the cost function of the form (\ref{eq:c_p}) with $p=l+1$ we can rewrite the estimate (\ref{eq:ubzeta}) for the difference in optimal objective values of problem (\ref{eq:multiperiod}) as
\begin{equation}
|v(P)-v(Q)|\le K\overset{\circ}{\mu}_{l+1}(P,Q)=K\hat\mu_{\hat c_{l+1}}(P,Q),
\label{eq:ubmu}
\end{equation}
where $K>0$ is some constant.

\section{\label{a:s-reduction}Scenario reduction algorithms}
The goal of scenario reduction algorithms is, given a stochastic optimization problem of the form (\ref{eq:gen_stoch}) characterized by a discrete measure $P=\sum_{i=1}^N p_i \delta_{\o_i}$ find the discrete measure $Q=\sum_{j=1}^M q_i \delta_{\tilde\o_j}$ such that $M<N$ and the difference in the optimal objective values $|v(P)-v(Q)|$ is as small as possible.

If the stochastic optimization problem has the form (\ref{eq:multiperiod}) of a linear multi-period problem then, as discussed earlier in this section, under relatively complete recourse assumption, the upper bound (\ref{eq:ubmu}) can be shown to hold. This motivates searching for discrete measures $Q$ that minimize the distance $\hat\mu_{l+1}(P,Q)$ (or $\overset{\circ}{\mu}_{l+1}(P,Q)$).

Thus the optimal scenario reduction problem based on the method of probability metrics can be formulated as follows \cite{dupacova2003}. Let $J\subset \{1,2,\dotsc, N\}$ and consider the measure $Q=\sum_{j\notin J} q_j\delta_{\o_j}$ supported at points $\o_j$, $j\in \{1,2,\dotsc, N\}\setminus J$. The measure $Q$ is said to be {\it reduced} from $P$ by deleting scenarios $\o_j$, $j\in J$ and by assigning new probabilities $q_j$ to the remaining scenarios. The optimal reduction concept proposed in \cite{dupacova2003} seeks the minimum value of the functional
\begin{equation}
D(J;q)=\hat\mu_p \left(\sum_{i=1}^N p_i\delta_{\o_i}, \sum_{j\notin J}q_j\delta_{\o_j} \right).
\label{eq:D(J:q)}
\end{equation}
It is shown in \cite{dupacova2003} that, for set $J$ fixed, the optimal weights $q$ are straightforward to find:
\begin{equation}
q_j=p_j+\sum_{i\in J_j} p_i,\;\; \text{for each}\; j\notin J,
\label{eq:redist}
\end{equation}
where $J_j := \{i\in J : j=j(i)\}$ and $j(i)\in \arg \min_{j\notin J} c_p(\o_i, \o_j)$ for each $i\in J$. The corresponding minimum of the functional $D(J;q)$ is
\begin{equation*}
D_J=\min_q \{D(J;q) : q_j\ge 0, \sum_{j\notin J} q_j =1 \}=\sum_{i\in J} p_i \min_{j\notin J}c_p (\o_i, \o_j).
\end{equation*}
On the other hand, the optimal choice of the set $J$ of given cardinality $|J|=k$
\begin{equation*}
\min_J \{D_J=\sum_{i\in J} p_i \min_{j\notin J}c_p (\o_i, \o_j) : J\subset \{1,2,\dotsc, N\}, |J|=k\}
\end{equation*}
is a combinatorial problem, and it is unlikely that efficient solution algorithms for arbitrary value of $k$ are available. However cases $k=1$ and $k=N-1$ are easy to solve to optimality and they can be used to formulate heuristic algorithms for other values of $k$. The {\it fast forward} scenario reduction algorithms proposed in \cite{heitsch2003} proceeds as follows.

{\bf Fast forward selection algorithm:}

{\bf Step 1:} $c_{ku}^{[1]} := c_p(\o_k, \o_u),\; k,u=1,\dotsc, N$,

$\displaystyle{z_u^{[1]} := \sum_{\substack{k=1 \\ k\ne u}} p_kc_{ku}^{[1]}},\; u=1,\dotsc N$,

$\displaystyle{u_1\in \arg \min_{u\in \{1,\dotsc, N\}} z_u^{[1]}},\; J^{[1]} := \{1, \dotsc, N\}\setminus \{u_1\}$.

{\bf Step $i$:} $c_{ku}^{[i]} := \min \{c_{ku}^{[i-1]}, c_{ku_{i-1}}^{[i-1]}\},\; k,u\in J^{[i-1]}$,

$\displaystyle{z_u^{[i]}:= \sum_{k\in J^{[i-1]}\setminus \{u\}} p_kc_{ku}^{[i]}},\; u\in J^{[i-1]}$,

$\displaystyle{u_i\in \arg \min_{u\in J^{[i-1]}}z_u^{[i]}},\; J^{[i]} := J^{[i-1]}\setminus \{ u_i \}$.

{\bf Step $n+1$:} Redistribution by (\ref{eq:redist}).



%

\end{document}